\documentclass[pra,twocolumn,superscriptaddress,notitlepage,amsmath,amssymb,showpacs,floatfix]{revtex4-1}
\usepackage{graphicx,subfigure,amssymb,amsmath,amsfonts}
\usepackage{verbatim}
\usepackage{ulem}
\usepackage{multirow}
\usepackage{hyperref}
\usepackage[T1]{fontenc} 

\newcommand{\be}{\begin{equation}}
\newcommand{\ee}{\end{equation}}
\newcommand{\ba}{\begin{eqnarray}}
\newcommand{\ea}{\end{eqnarray}}
\newcommand{\njp}{New J. Phys.}
\newcommand{\prx}{Phys. Rev. X}

\usepackage{dsfont}
\usepackage[usenames]{color}

\hypersetup{
    colorlinks=true,       		
    linkcolor=black,          
    citecolor=black,          
    filecolor=black,      		
    urlcolor=black,           
    runcolor=cyan
}

\DeclareSymbolFont{AMSb}{U}{msb}{m}{n}
\DeclareSymbolFontAlphabet{\mathbb}{AMSb}

\begin{document}
\title{Two-dimensional network of atomtronic qubits}

\author{S. Safaei} 
\affiliation{Centre for Quantum Technologies, National University of Singapore, 3 Science Drive 2, 117543 Singapore}
\affiliation{MajuLab, CNRS-UCA-SU-NUS-NTU International Joint Research Unit, Singapore} 

\author{B. Gr\'{e}maud}
\affiliation{MajuLab, CNRS-UCA-SU-NUS-NTU International Joint Research Unit, Singapore}
\affiliation{Centre for Quantum Technologies, National University of Singapore, 3 Science Drive 2, 117543 Singapore}
\affiliation{Physics Department, Faculty of Science, National University of Singapore, 2 Science Drive 3, 117551 Singapore}
\affiliation{Laboratoire Kastler Brossel, Ecole Normale Sup\'{e}rieure CNRS, UPMC; 4 Place Jussieu 75005 Paris, France}

\author {R. Dumke} 
\affiliation{Centre for Quantum Technologies, National University of Singapore, 3 Science Drive 2, 117543 Singapore}
\affiliation{Division of Physics and Applied Physics, School of Physical and Mathematical Sciences, Nanyang Technological University, 21 Nanyang Link, Singapore 637371}
\affiliation{MajuLab, CNRS-UCA-SU-NUS-NTU International Joint Research Unit, Singapore}

\author{L.-C. Kwek} 
\affiliation{Centre for Quantum Technologies, National University of Singapore, 3 Science Drive 2, 117543 Singapore}
\affiliation{National Institute of Education and Institute of Advanced Studies, Nanyang Technological University, 1 Nanyang Walk, Singapore 637616}
\affiliation{Institute of Advanced Studies, Nanyang Technological University, 60 Nanyang View, Singapore 639673, Singapore}
\affiliation{MajuLab, CNRS-UCA-SU-NUS-NTU International Joint Research Unit, Singapore}

\author{L. Amico}
\affiliation{INFN-Laboratori Nazionali del Sud, INFN, via S. Sofia 62, 95123 Catania, Italy.}
\affiliation{ CNR-MATIS-IMM $\&$ Dipartimento di Fisica e Astronomia,   Via S. Sofia 64, 95127 Catania, Italy}
\affiliation{Centre for Quantum Technologies, National University of Singapore, 3 Science Drive 2, 117543 Singapore}

\author{C. Miniatura}
\affiliation{MajuLab, CNRS-UCA-SU-NUS-NTU International Joint Research Unit, Singapore}
\affiliation{Centre for Quantum Technologies, National University of Singapore, 3 Science Drive 2, 117543 Singapore}
\affiliation{Physics Department, Faculty of Science, National University of Singapore, 2 Science Drive 3, 117551 Singapore}
\affiliation{Division of Physics and Applied Physics, School of Physical and Mathematical Sciences, Nanyang Technological University, 21 Nanyang Link, Singapore 637371}
\affiliation{Universit\'e C\^ote d'Azur, CNRS, INPHYNI; France}
\affiliation{Institute of Advanced Studies, Nanyang Technological University, 60 Nanyang View, Singapore 639673, Singapore}

\begin{abstract}
Through a combination of laser beams, we engineer a $2$D optical lattice of Mexican hat potentials able to host atoms in its ring-shaped wells. When tunneling can be ignored (at high laser intensities), we show that a well-defined qubit can be associated with the states of the atoms trapped in each of the rings. Each of these two-level systems can be manipulated by a suitable configuration of Raman laser beams imprinting a synthetic flux onto each Mexican hat cell of the lattice. Overall, we believe that the system has the potential to form a scalable architecture for atomtronic flux qubits.
\end{abstract}


\maketitle

\section{Introduction}
\label{intro}
Atomtronics aims at exploiting the matter wave aspects of quantum 
cold-atom systems confined in magnetic or laser light circuits of complex 
shapes~\cite{Seaman,amico2015atomtronics,AtomtronicsFocusIssue}. Several 
elementary atomtronic devices and circuits have already been 
proposed \cite{micheli2004, stickney2007, pepino2009, pepino2010} and some have been 
realized~\cite{caliga2015,ramanathan,eckel2014,ryu2007,Ryu2015}. 
The construction of atomtronic integrated circuits, though, remains an important 
open problem not only in quantum optics but in the broader field of quantum technology. 
In this paper, we propose a scheme to create a network of atomtronic rings with the potential 
to be used as flux qubits for information processing. Crucially, the approach might prove scalable.

Qubits can be implemented in a variety of physical systems~\cite{superconducting,coldatoms,rydberg,iontraps, NMR,qdots} with different advantages and disadvantages. Solid-state realizations allow the 
construction of fast gates (nanoseconds) but need to operate at short time scales (microseconds) 
to fight decoherence and/or dissipation. An important advantage of such configurations is that 
they benefit from the scalability provided by highly-developed lithographic techniques. On the 
other hand, atomic qubits realized by hyperfine states of cold atoms confined in optical lattices 
have very long storage and coherence times (fraction of a second). For such systems, scalability 
has been achieved in principle~\cite{litho_Dumke}, but single-site addressability is the main 
bottleneck in quantum processing with cold atoms.
 
With atomtronic flux qubits, we seek to combine the macroscopic quantum 
coherence of the Josephson junction flux qubits with the advantages of cold 
atoms~\cite{solenov_qubit,Amico_qubit,aghamalyan2015atomtronic}. The devices 
have the phenomenology of an atomtronic quantum interference device (AQUID), 
the atomic counterpart of a superconducting quantum interference device 
(SQUID), and they operate with a ring-shaped Bose-Einstein condensate (BEC). 
The two-level system is based on clockwise and anti-clockwise atomic currents 
obtained by applying an effective gauge field to the system~\cite{dalibard}. 
In the simplest scheme, superpositions of these current states are generated 
by forward and back scattering flows of the cold atoms through a single tunnel 
barrier (weak link) that is imprinted along the ring-shaped potential (breaking 
the Galilean invariance of the system). Although schemes for single or few 
coupled atomtronic qubits have been conceived~\cite{Amico_qubit,couple_rings}, 
the implementations are complex. As a consequence, it is challenging to take 
a "bottom-up" approach to a scalable architecture. Instead, in this paper, we 
pursue a "top-down" approach.

We propose a laser scheme to realize a pattern of closed 
circular currents arranged in a planar configuration. 
Such a pattern emerges from a two-dimensional (2D) optical lattice consisting of a triangular 
periodic array of Mexican hat potentials. Atoms can be trapped in its 
{\it{nearly}} ring-shaped confining wells. The scheme is completed by applying 
a suitable laser configuration subjecting the lattice to an effective gauge field. 
We demonstrate that an effective two-level system arises in each elementary cell 
of the 2D lattice. Furthermore, the system can be controlled by the effective 
gauge field. Overall, our system would potentially constitute a 2D architecture 
hosting flux qubits. We mention possible schemes to address, couple, 
and manipulate the two-level systems arranged in such a 2D Mexican hat lattice.

The rest of the paper is organized as follows. In Sec.~\ref{potential} we explain 
the laser configuration used to produce the lattice of Mexican hat potentials. 
In Sec.~\ref{band} we discuss the condition under which the Mexican hats (rings) 
are practically decoupled. Next, in Sec.~\ref{qubit}, we show how different parameters 
of the system can be tuned in order to engineer the energy spectrum of each single Mexican 
hat and obtain a spectrum similar to the one of superconducting flux qubits. We discuss the 
feasibility of the system, referring to typical required experimental parameters, in Sec.~\ref{feasibility} 
and briefly mention how qubit gates could be implemented in Sec.\ref{gates}. 
We summarize our work and conclude with some perspectives in Sec.~\ref{sum}.

\section{2D Mexican hat lattice laser configuration}
\label{potential}
We consider atoms (mass $m$, resonance frequency $\omega_{at}$, linewidth $\Gamma$) subjected to three coplanar standing waves lying in the $xy$-plane at relative angles $\pi/3$ to each other. They are produced by three retro-reflected monochromatic laser beams (same frequency $\omega_L$) linearly-polarized along axis $Oz$. The corresponding wave vectors are 
$\vec{k}_1=k_L\Big(\frac{\sqrt{3}}{2}\hat{x}+\frac{1}{2}\hat{y}\Big)$, 
$\vec{k}_2=k_L\Big(-\frac{\sqrt{3}}{2}\hat{x}+\frac{1}{2}\hat{y}\Big)$, and 
$\vec{k}_3=\vec{k}_1+\vec{k}_2=k_L\hat{y}$, 
with $k_L=\omega_L/c=2\pi/\lambda_L$ ($\lambda_L$ is the laser wavelength), and we assume their respective Rabi frequencies to be $\Omega_1=\Omega_2=\gamma \Omega$ and $\Omega_3=\Omega$. The externally adjustable parameter $\gamma$ is the relative strength of the two lateral standing waves compared to the one along $Oy$. For far blue-detuned laser beams (positive detuning $\delta_L = \omega_L-\omega_{at} \gg \Gamma$), and after a suitable choice of the origin of coordinates, the light-shift potential experienced by the atoms is $V(\vec{r}) = U_0 v(\vec{r})$ where:

\begin{eqnarray}
\label{eq:veff2}  
v(\vec{r})=\Big[\cos k_Ly+2\gamma\cos(\frac{k_Ly+\phi}{2})\cos(\frac{\sqrt{3}k_Lx}{2})\Big]^2
\end{eqnarray}
and $U_0=\hbar\Omega^2/(4\delta_L)>0$. 
Note that fixing the origin imposes two conditions on the phases of the lasers, 
thus leaving only one adjustable phase parameter, $\phi$, in the equation above.
The full optical potential shows up as a triangular lattice of Mexican hat structures, with the unit Bravais cell being spanned by 
$\vec{a}_1=\lambda_L(\frac{1}{\sqrt{3}}\hat{x}+\hat{y})$ and 
$\vec{a}_2=\lambda_L(-\frac{1}{\sqrt{3}}\hat{x}+\hat{y})$~\cite{safaei_2015}.

\begin{figure}
\includegraphics[width=0.4\textwidth]{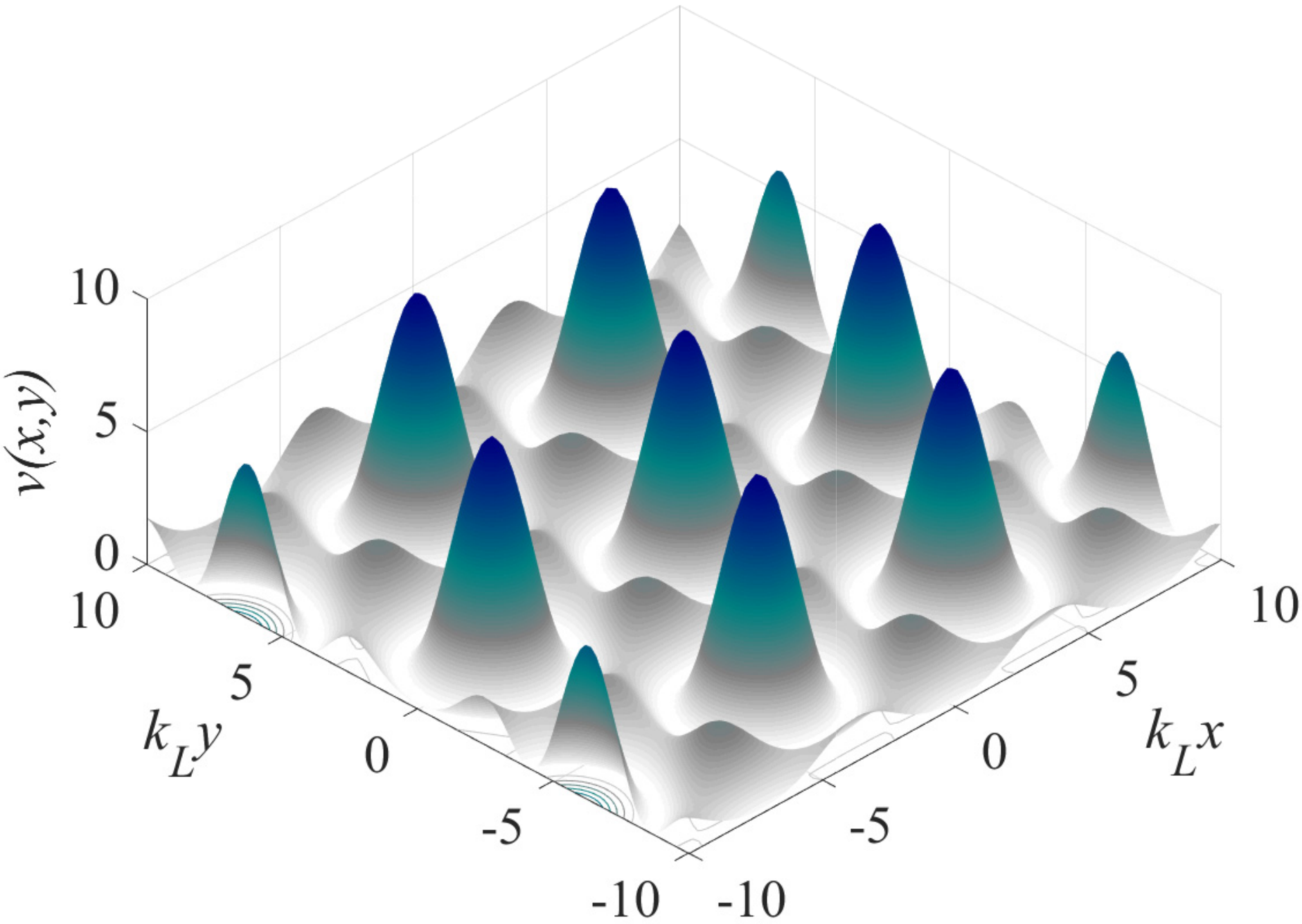}
\includegraphics[width=0.3\textwidth]{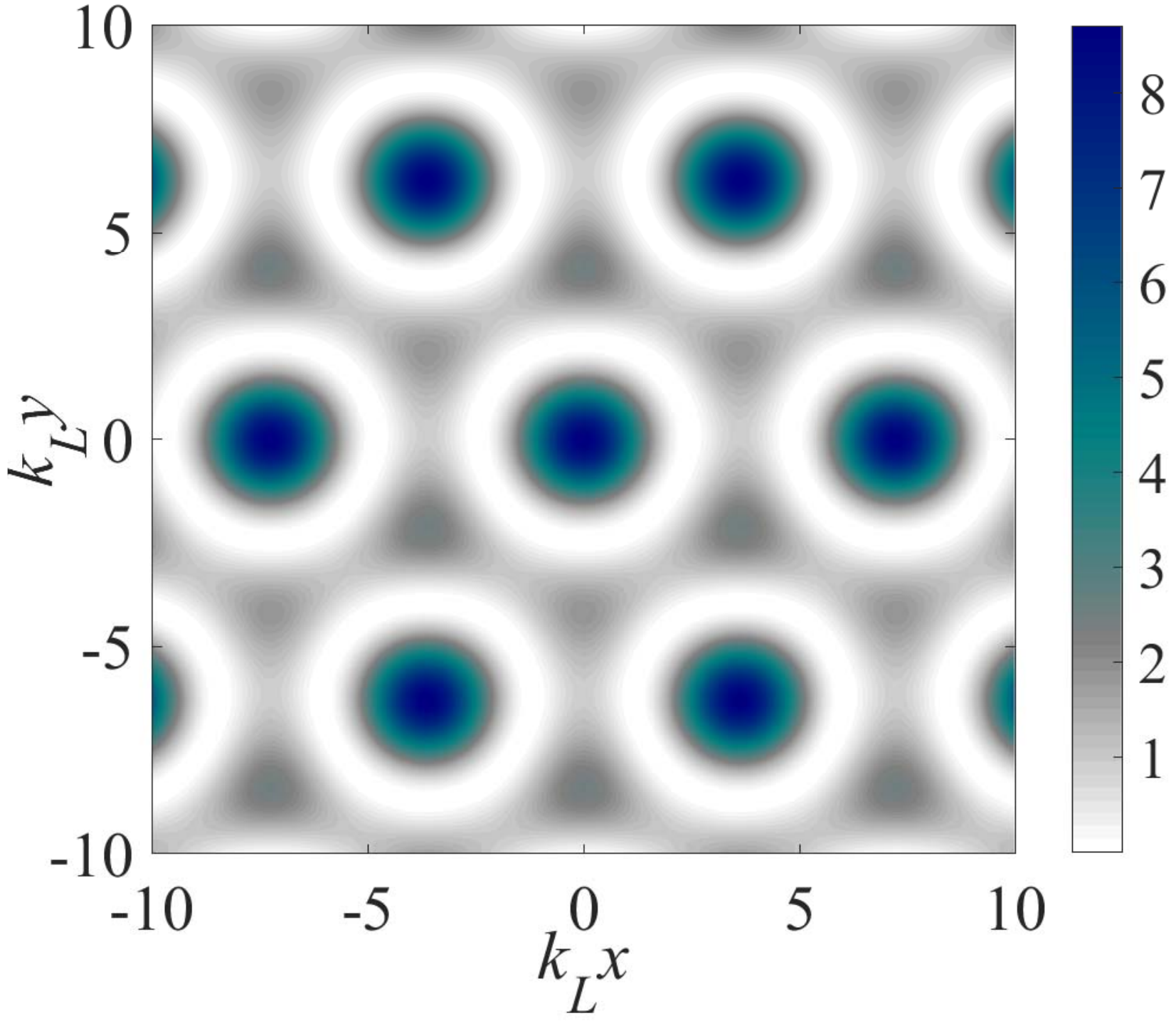}
\includegraphics[width=0.3\textwidth]{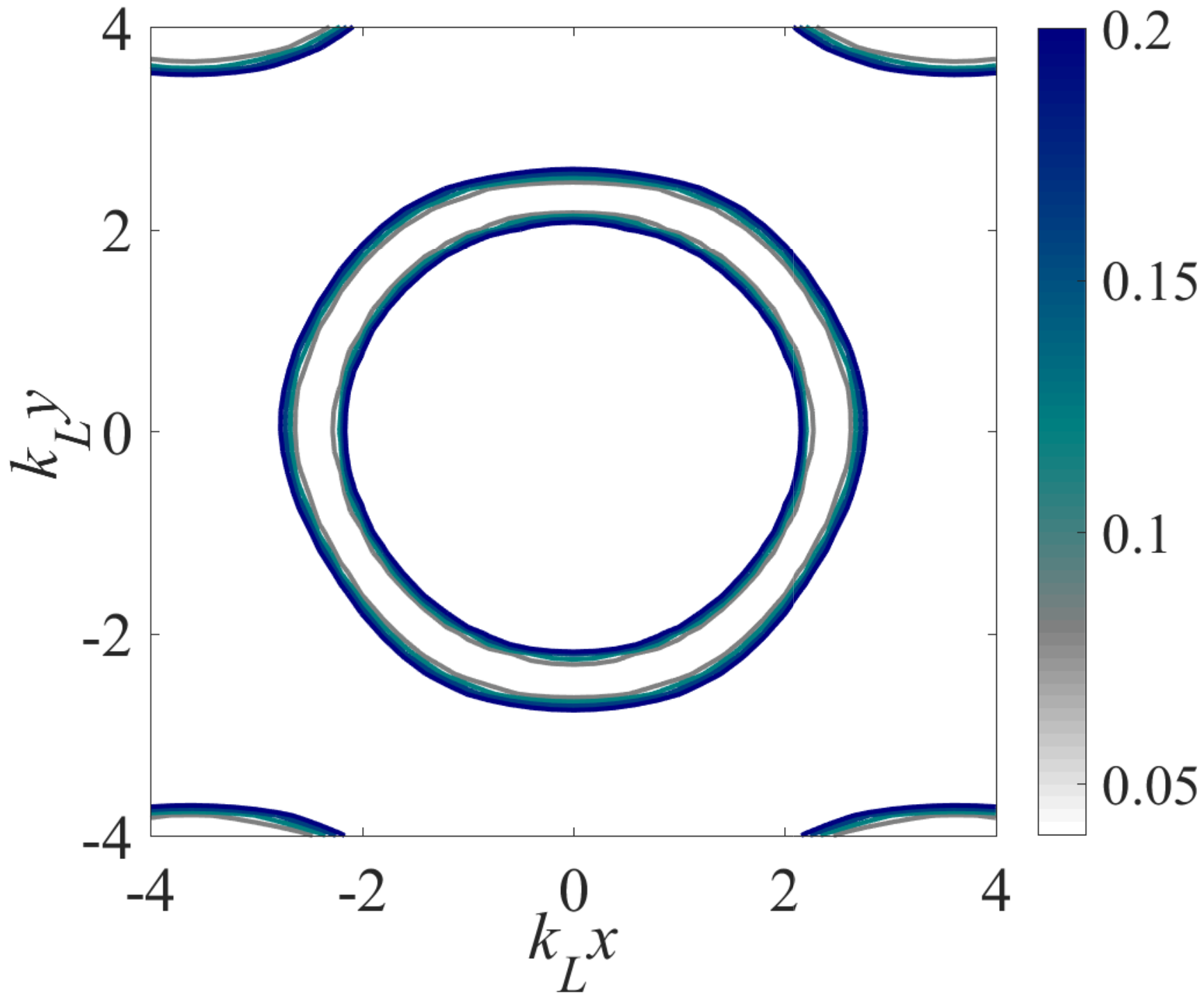}
\caption{(Color online) The triangular optical lattice of Mexican hats obtained from the structure function 
$v(\vec{r})$, Eq.~(\ref{eq:veff2}), with $\gamma=0.98$ and $\phi=\pi/25$ (top panel). Sufficiently cold atoms 
would accumulate in the ring-shaped minima obtained for $v(\vec{r})=0$ (white rings in the middle panel). By 
increasing the potential strength $U_0$, tunneling between adjacent cells can be strongly suppressed and the 
different cells become independent. Each of them is able to store a single "flux" qubit. Bottom panel: Contour 
plot of the slightly distorted ring-shaped potential well within a unit cell of the Mexican hat lattice.
\label{fig:lattice}}
\end{figure} 

This Mexican hat structure is slightly distorted but is maintained provided the lattice laser beams 
are not too imbalanced ($\gamma$ sufficiently close to unity) and almost in-phase ($\phi$ small enough). 
Figure~\ref{fig:lattice} gives a plot of $v(\vec{r})$ and of the ring structure of its minima for 
$\gamma=0.98$ and $\phi=\pi/25$. 

\section{Independent lattice cells regime}
\label{band}
\begin{figure}
\begin{center}
\includegraphics[width=0.48\textwidth]{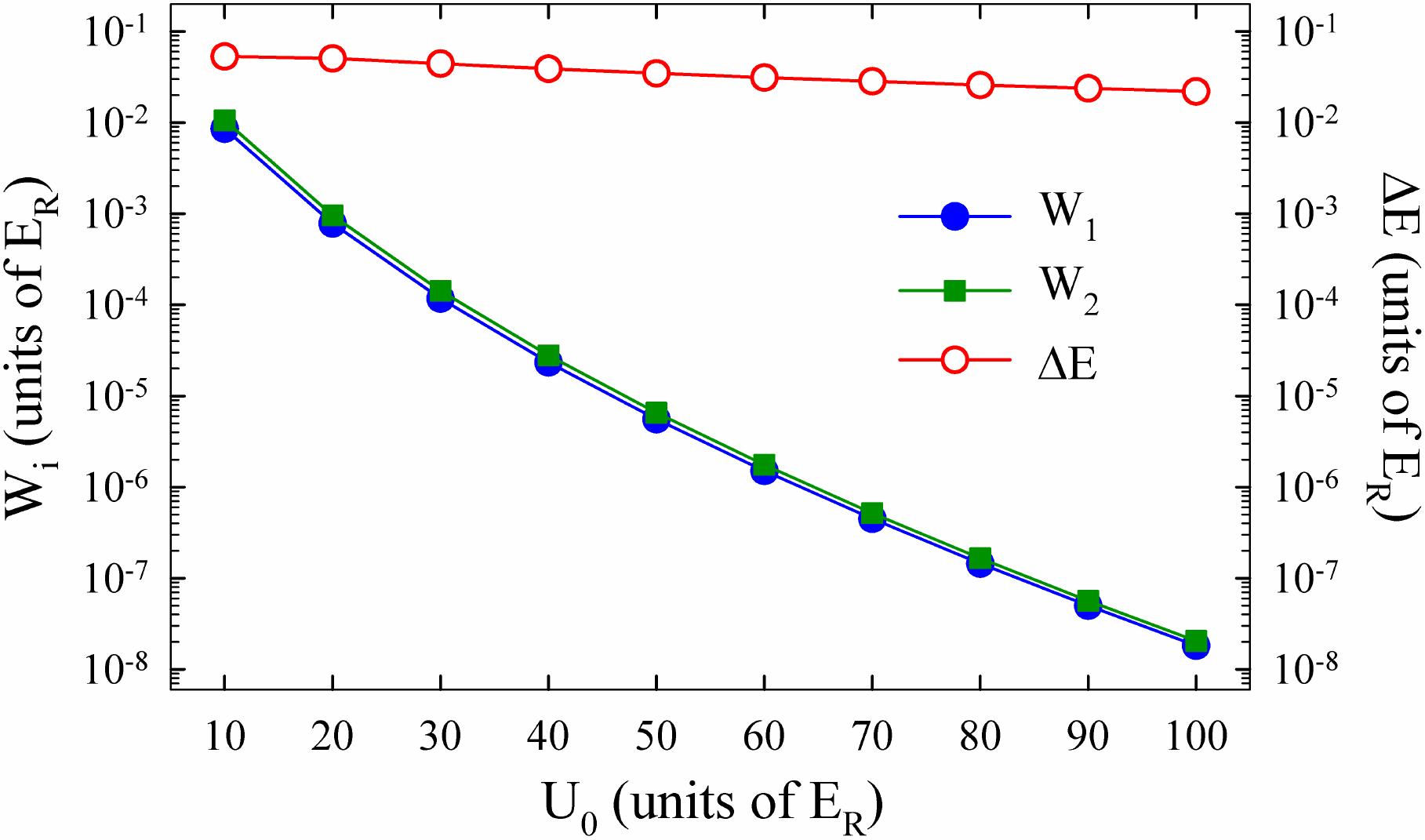}
\end{center}
\caption{(Color online) 
Logarithmic plot of the gap $\Delta E$ (in units of the recoil energy $E_R$) between the two lowest bands of the Mexican hat lattice and their respective widths $W_1$ and $W_2$ (in units of $E_R$) for different values of $U_0/E_R$. All plots are obtained with $\gamma=0.98$ and $\phi=\pi/25$. For $U_0 \geq 50E_R$, the band widths are smaller than the band gap by at least four orders of magnitude. 
From a numerical fit we find the value of the effective action $S$ to be around $3.4$ (see text).
\label{fig:bands}}
\end{figure}

When the lower bands of the Mexican hat lattice band structure are flat compared to their separation, tunneling 
does not efficiently couple adjacent cells. This means that atoms trapped in a given cell would stay there for a 
very long time and would be virtually isolated from the rest of the lattice. Providing this residence time (given 
by the tunneling time) is larger than the time required to manipulate and interrogate the atoms, then their local 
dynamics can be simply understood in the so-called atomic limit, that is from the local eigenstates and spectrum 
of the Mexican hat potential within one cell. The tunneling amplitude between adjacent cells is expected to scale 
as $\hbar_e^{-3/2}\exp(-S/\hbar_e)$, where $S$ is a number (effective action) and $\hbar_e=\sqrt{2E_R/U_0}$ is the 
effective Planck's constant ($E_R= \hbar^2k_L^2/(2m)$ is the recoil energy)~\cite{Lee09}. Therefore inter-cell tunneling is exponentially suppressed with a rate proportional to $\sqrt{U_0/(2E_R)}$. At the same time, the band 
gap is expected to scale algebraically with $\hbar_e$ (the power law depends on the anharmonicity of the potential 
around its minimum). Thus, the larger the $U_0$ the flatter are the bands and the better is the ratio between the 
band gap and the band-width. Figure\ref{fig:bands} shows our data extracted from a numerical computation of the 
band structure. As one can see, for $U_0\geq50E_R$, the band widths are smaller than the band gap by more than 
four orders of magnitude.

\section{Local qubit system}
\label{qubit}
The basic idea  is to associate a qubit with the states of the atoms confined within 
the unit cell of the lattice of Mexican hats. Note that here, local rotation and Galilean 
invariance is broken by the distortions of the ring-shaped wells (see the bottom panel of 
Fig.\ref{fig:lattice}). Control of the level splitting can be achieved by imparting a 
synthetic flux through Laguerre-Gauss beams and two-photon stimulated Raman processes to 
transfer orbital angular momentum to the atoms \cite{PhysRevLett.97.170406,Phase_imprinting_Adiabatic_Passage}. 
Many-cell addressing can be done by using optical vortex arrays \cite{Lin2011}, or by using 
a hologram generated by a spatial light modulator (SLM) \cite{Gauthier2016}, while individual 
addressing can be achieved by using a high-resolution objective and a $XY$-scanning acousto-optic 
modulator (AOM) configuration \cite{Labuhn2014}. 

The single-cell and single-particle Hamiltonian for this system reads

\begin{equation}
\mathcal{H} = \frac{(\vec{p}-\vec{A})^2}{2m} + U_0 v(\vec{r})
\end{equation}

where $\vec{r}=\alpha_1 \vec{a}_1+\alpha_2 \vec{a}_2$ is restricted within a given unit Bravais cell $\mathcal{B}$ ($|\alpha_i| \leq 1/2$, $i=1,2$) of the full lattice and where open boundary conditions are used ($\psi(\vec{r}) = 0$ for $\vec{r}\in \partial\mathcal{B}$). The synthetic gauge field can be chosen as
$\vec{A}=- By\hat{x}$, providing an effective  magnetic field $B$ along $Oz$ and a flux per unit cell $\Phi=(\vec{\nabla}\times\vec{A}).(\vec{a}_1\times\vec{a}_2)=2B\lambda_L^2/\sqrt{3}$. We have checked that the lowest eigenenergies of this system at zero flux match with the ones obtained from the band structure of the full lattice at zero Bloch wave vector.

\begin{figure}
\includegraphics[width=0.32\textwidth]{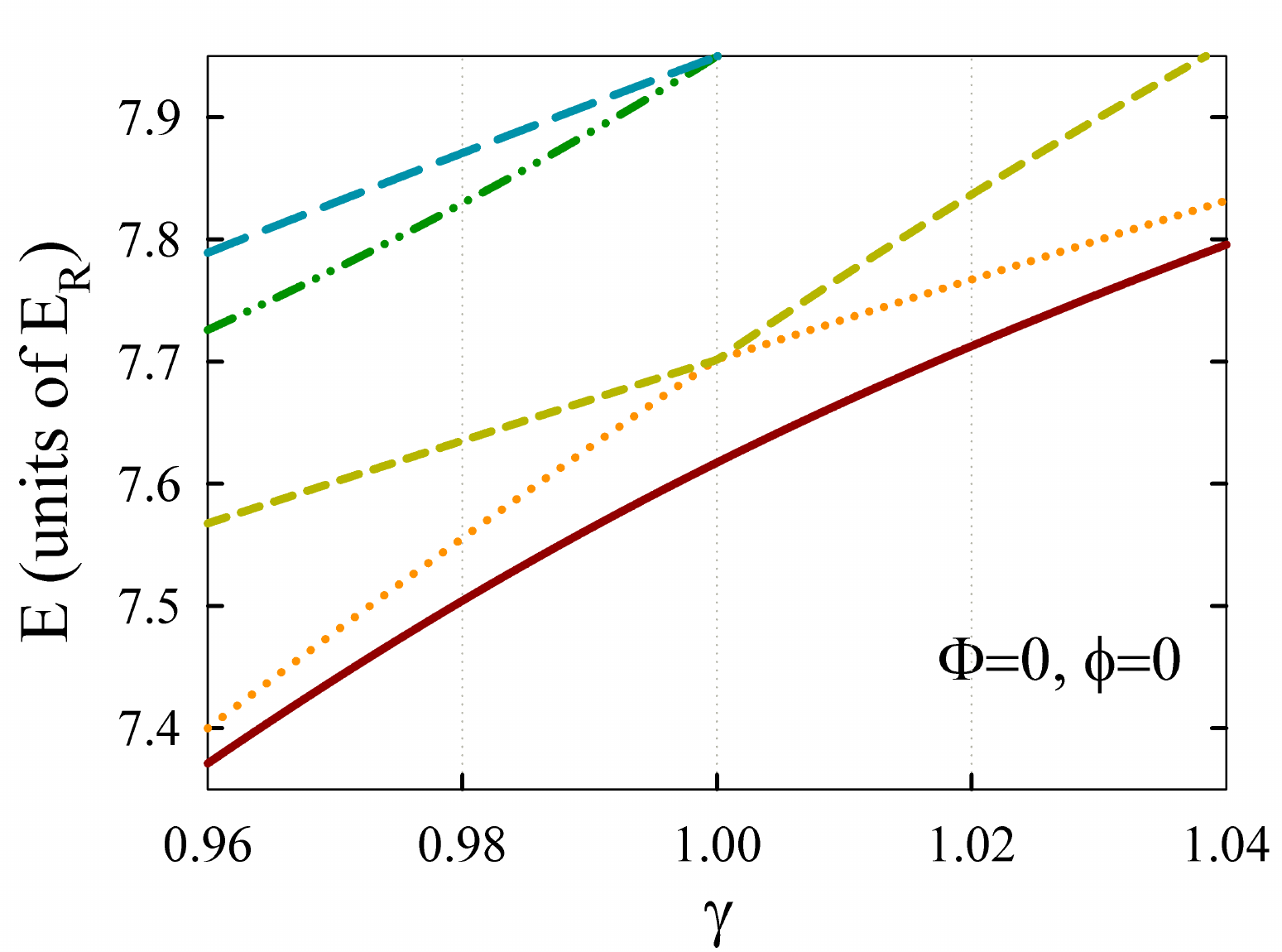}\hfill
\includegraphics[width=0.32\textwidth]{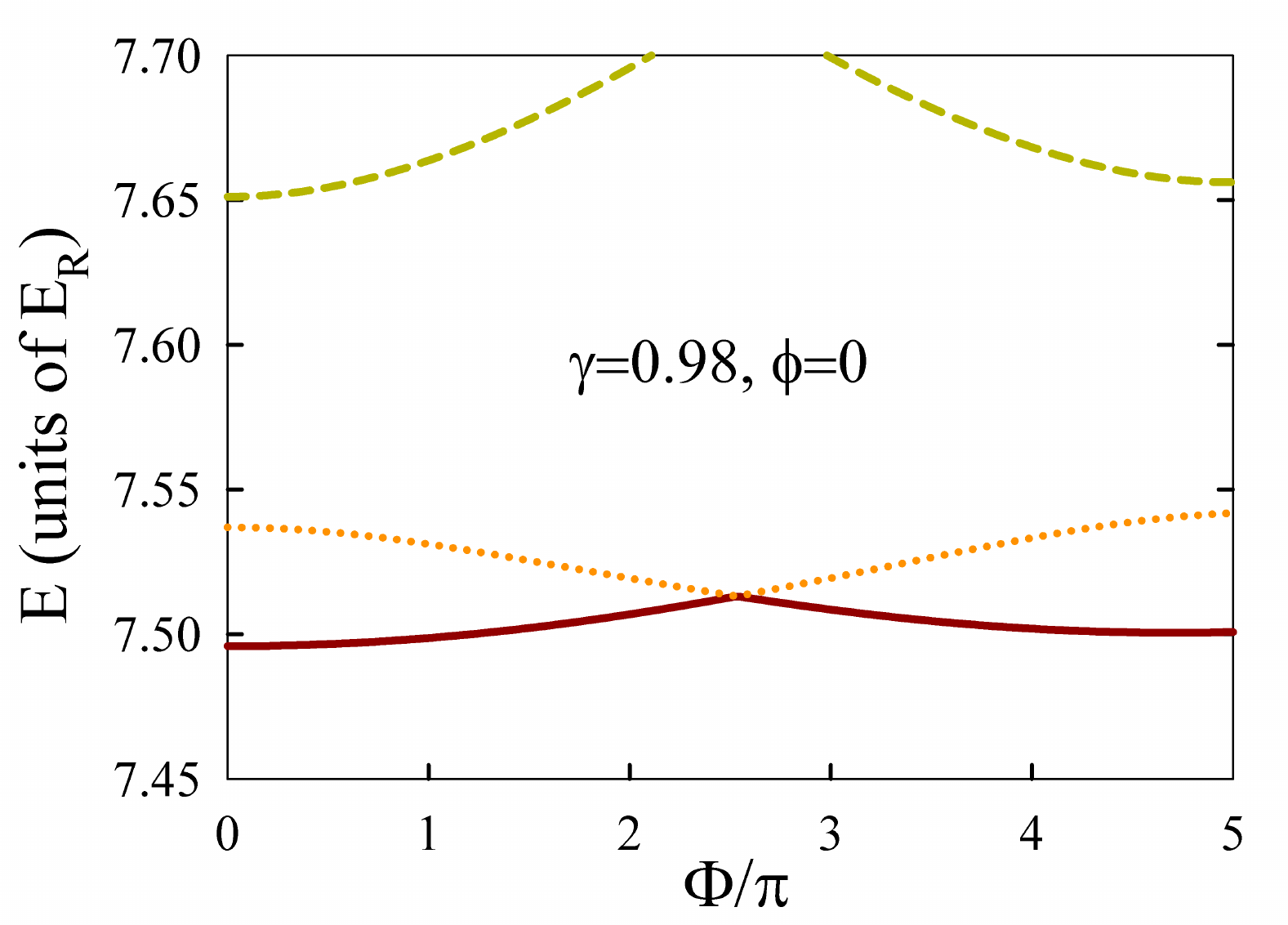}\\
\vspace{0.2cm}
\includegraphics[width=0.32\textwidth]{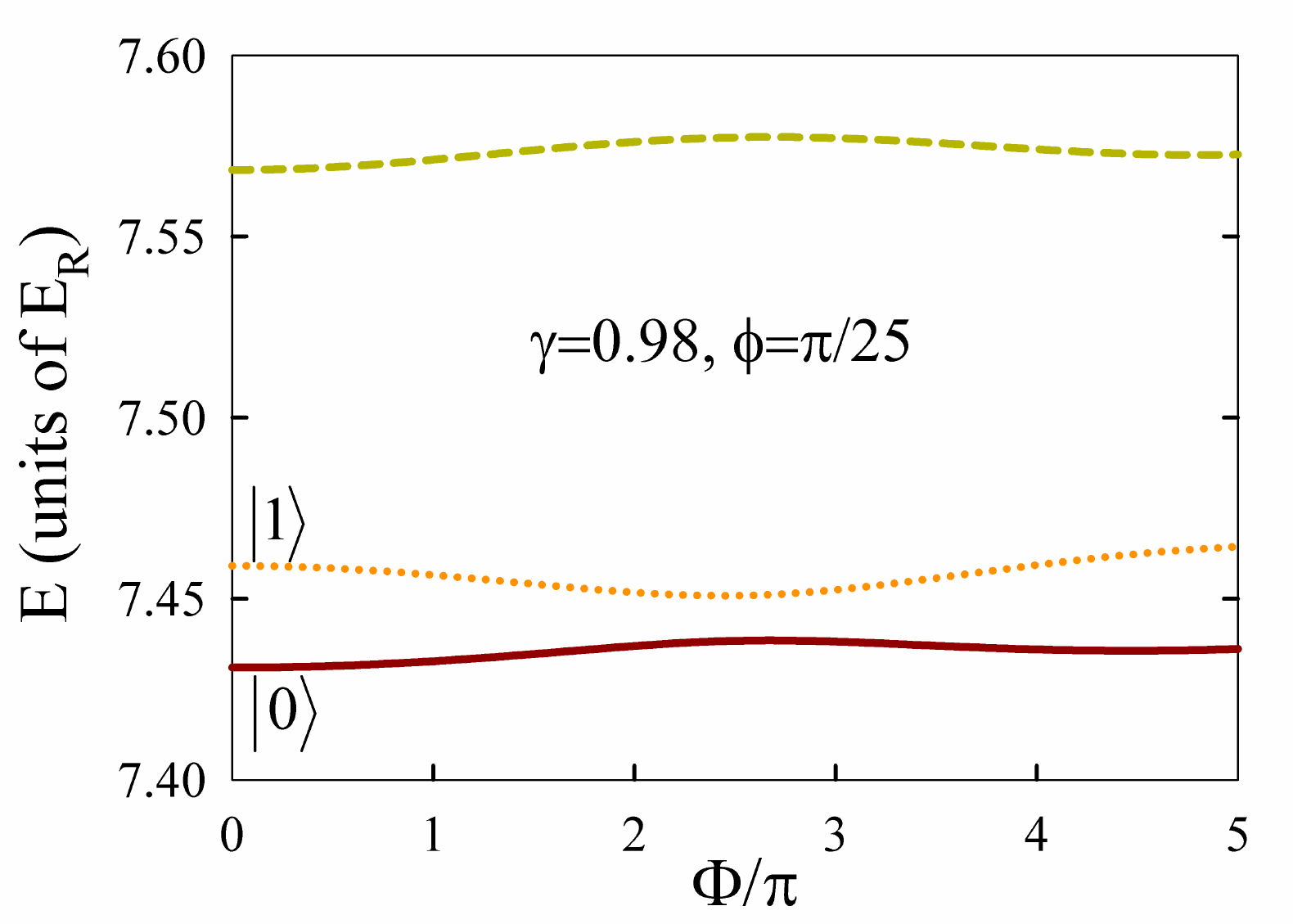}\hfill
\caption{(Color online) Top panel: The five lowest zero-flux energy levels of the lattice band structure at zero Bloch wave vector (in units of the recoil energy $E_R$) as a function of imbalance $\gamma$ when $\phi=0$. By departing $\gamma$ from unity, one can separate the two lowest levels from all others. Middle panel: The three lowest single-particle energy levels (in units of the recoil energy $E_R$) obtained for $\gamma=0.98$ and $\phi=0$ within a single unit Bravais cell (with open boundary conditions) as a function of the synthetic flux $\Phi$. The two lowest levels cross at some flux $\Phi_0=2.525\pi$. Bottom panel: Same as the middle panel but with 
$\phi=\pi/25$. As one can see, the degeneracy at $\Phi_0$ is lifted. The qubit is encoded in the two lowest states dubbed $|0\rangle$ and $|1\rangle$. For all panels, $U_0=50E_R$. 
\label{fig:E2}}
\end{figure}

Figure~\ref{fig:E2} shows the behavior of the lowest single-particle energy levels as functions of parameters $\gamma$, $\phi$ and flux $\Phi$. Starting from the lattice band structure obtained at zero Bloch wave vector, we isolate two levels from all others by departing $\gamma$ from unity (top panel). Choosing $\gamma=0.98$, we next compute how these levels change with the flux $\Phi$ generated by an artificial gauge field imprinted on the atoms (middle panel). The two lowest levels cross at some flux $\Phi_0\approx 2.525\pi$. A small  phase difference $\phi=\pi/25$ then serves to lift the degeneracy at $\Phi_0$, the third level being still sufficiently away (bottom panel). For this set of parameters, we thus get the typical level behavior of flux qubits with an avoided crossing. We use the corresponding states, dubbed $|0\rangle$ and $|1\rangle$, to encode a qubit in each of the unit cells of the lattice of Mexican hats. 

\section{Feasibility}
\label{feasibility}
Considering, as an example, $^{87}Rb$ atoms and the $S_{1/2}$ - $P_{3/2}$ transition 
($\omega_{at}=2\pi\times384.23$ THz, $\lambda_{at}=780.24$ nm, $\Gamma=2\pi\times6.06$ MHz), 
one could use blue-detuned lasers by $\delta_L= 2\pi\times6$ THz 
(or, equivalently, $\lambda_{at}-\lambda_L=12$ nm) to produce the lattice. In this case, 
a lattice with overall strength $U_0=100 E_R$ requires laser intensities $I=4.1$ MW/m$^2$. 
Therefore, lasers with a power of $1$W would be able to produce a lattice area of 
$500 \mu m\times500 \mu m$ which would host more than $250000$ unit-cells/qubits. Stabilizing 
the lattice strength $U_0$ at a level of $4\%$ is achieved by stabilizing the laser output power 
at the same level (or at $20\%$ in Rabi frequency), which is feasible. The lattice structure is 
determined by the values chosen for $\phi$ and $\gamma$. Taking $\phi=0$ as the reference point, 
setting $\phi = \pi/25 = 0.02 \times 2\pi$ requires moving the mirror along $Oy$ by $0.02~\lambda_L =15.4$ nm. 
This is within the range of the current technology which allows precise and stable nanometer 
positioning \cite{Schmiegelow2016,Teo2015}. Fixing $\gamma=0.98$ requires fixing the ratio of the 
Rabi frequencies with a precision better than $2\%$.

\section{Comments on  quantum gates and read-out}
\label{gates}
\begin{figure}
\includegraphics[width=0.23\textwidth]{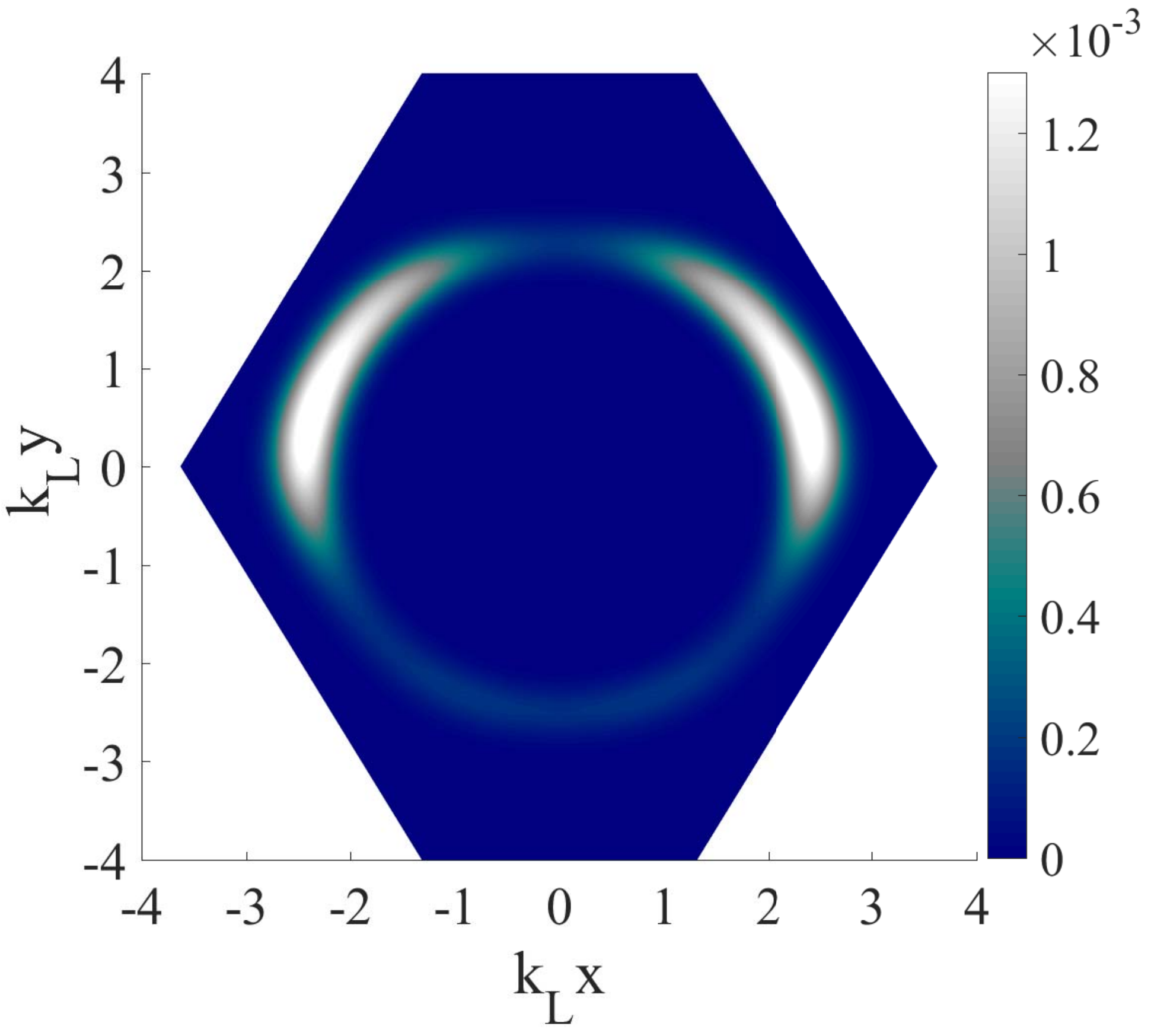}\hfill
\includegraphics[width=0.23\textwidth]{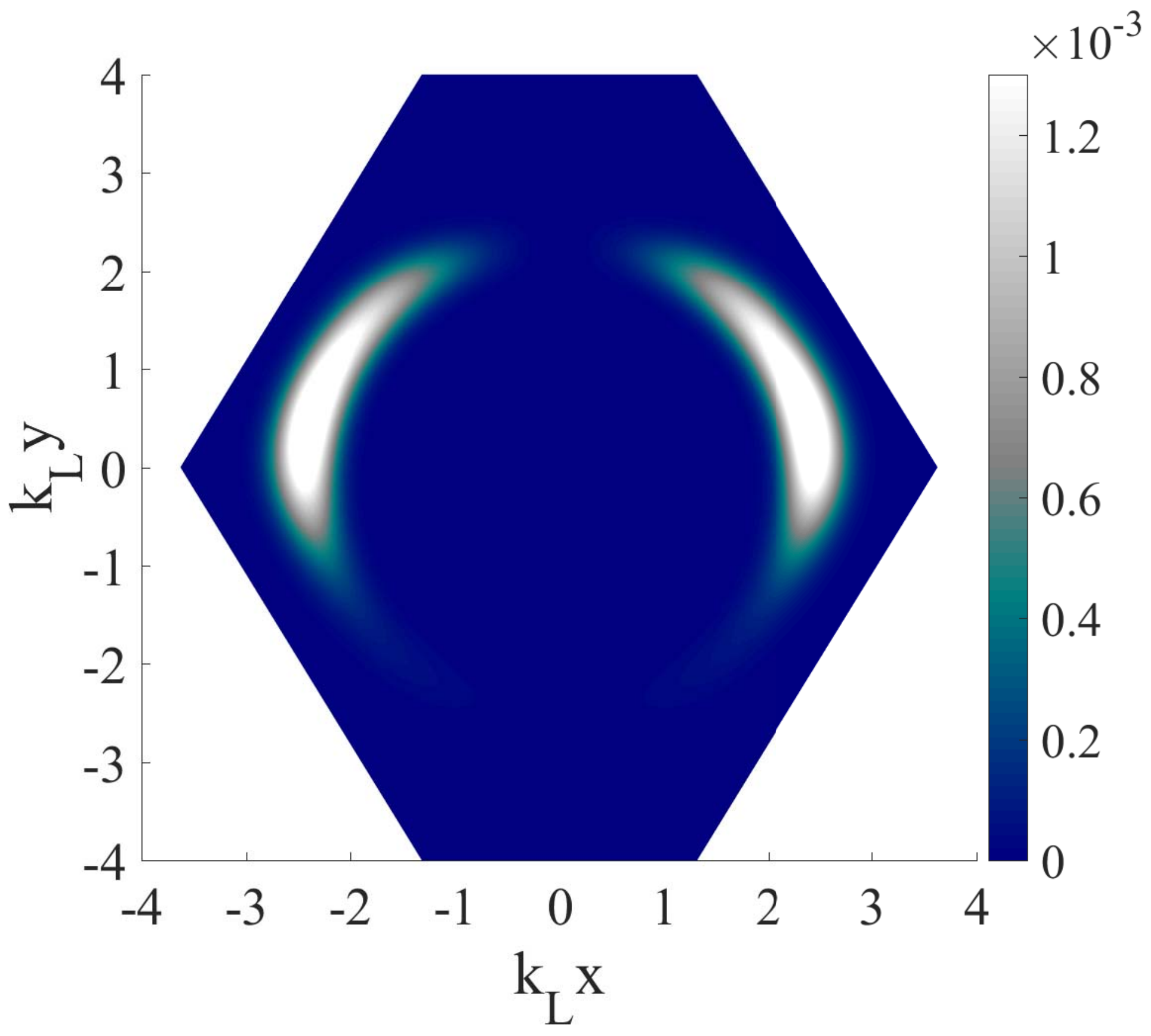}\\
\includegraphics[width=0.23\textwidth]{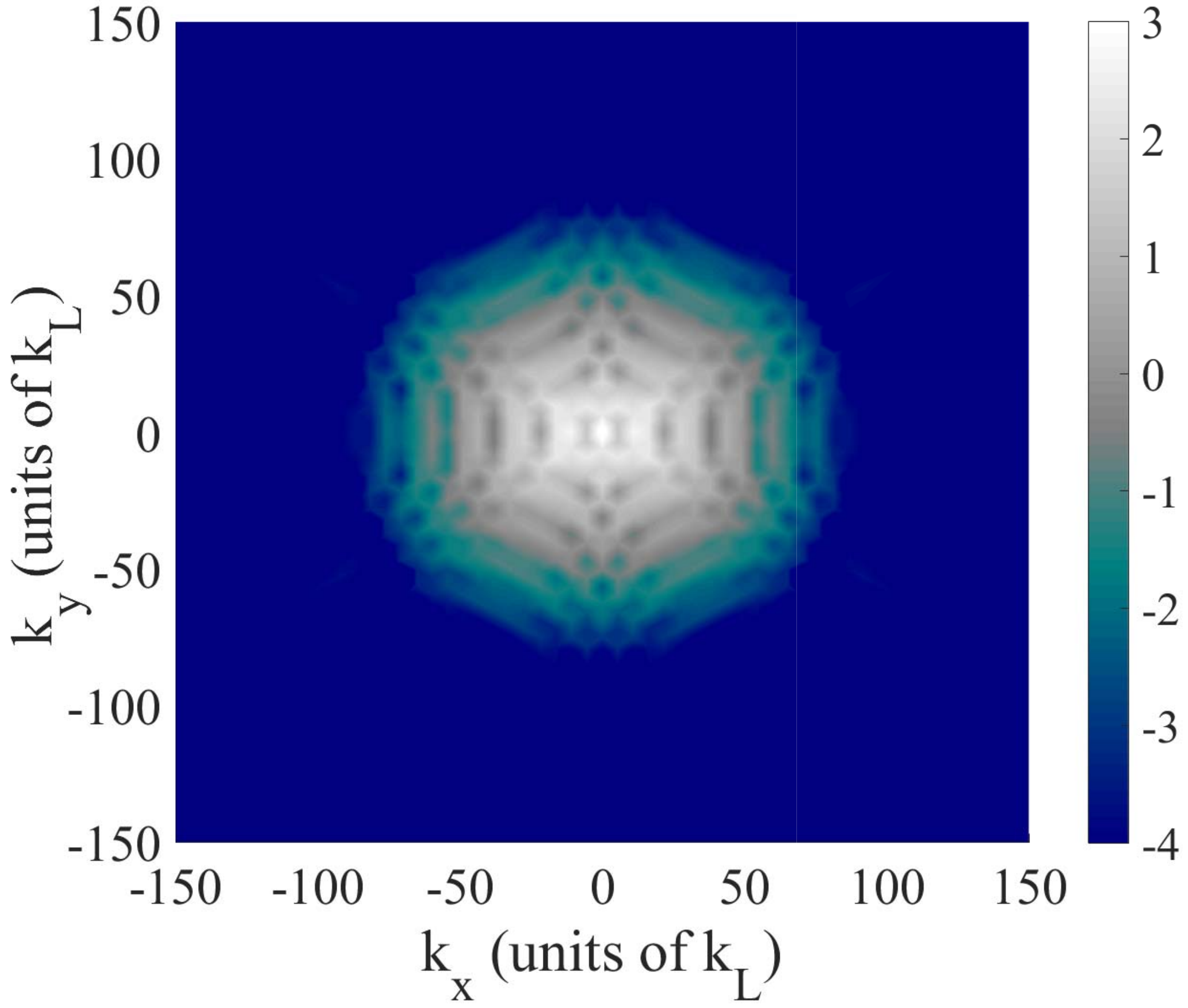}\hfill
\includegraphics[width=0.23\textwidth]{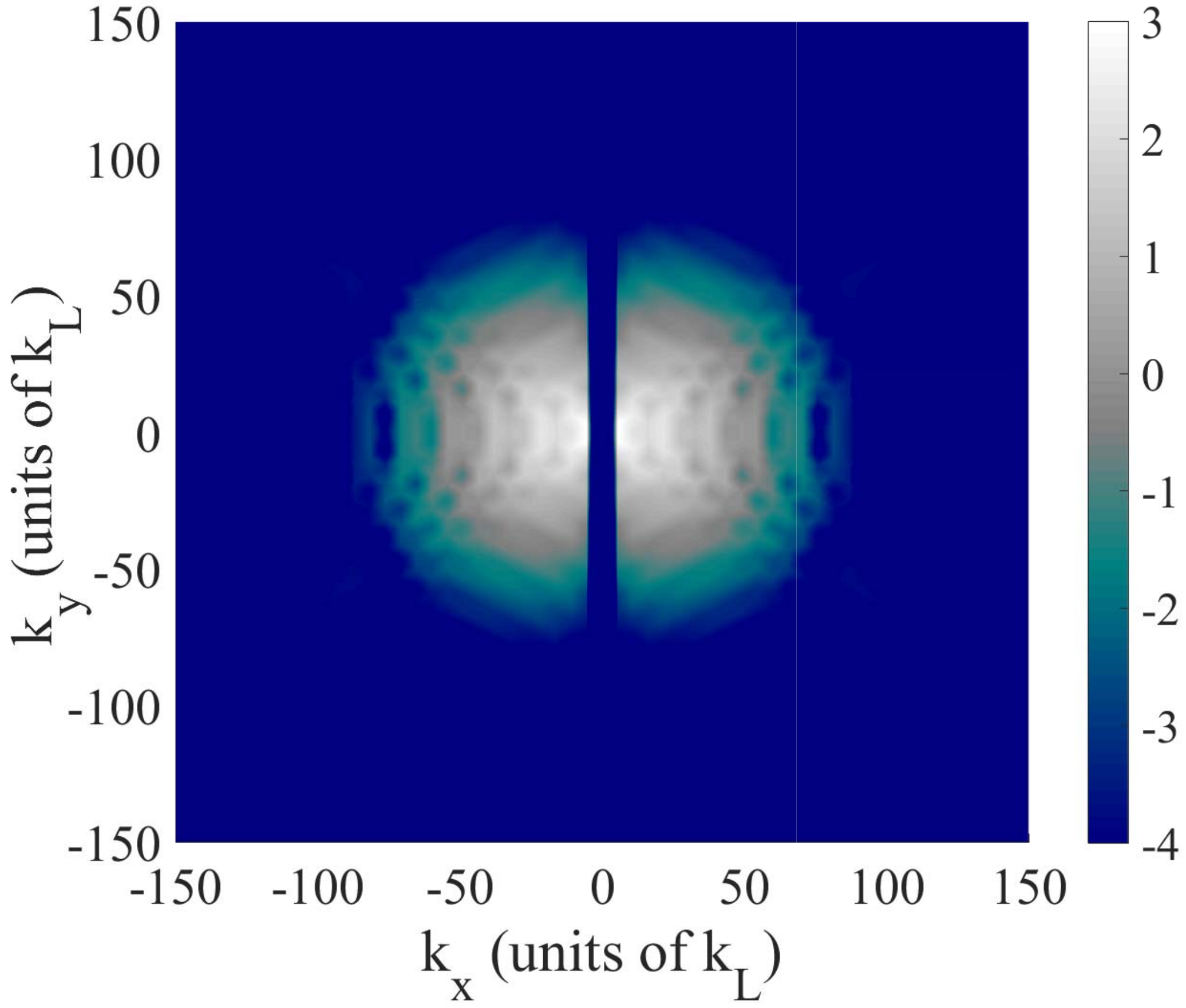}
\caption{(Color online) 
Top panels: Spatial density distributions, $|\psi(\vec{r})|^2$, of qubit states $|0\rangle$ (left) and 
$|1\rangle$ (right) at zero flux $\Phi=0$. 
Bottom panels: Logarithm of the momentum density distributions, $\log(|\psi(\vec{k})|^2)$, of the same states at zero flux. Since states $|0\rangle$ and $|1\rangle$ are 
respectively even and odd with respect to $x\to-x$, so are their Fourier transforms with respect to $k_x\to-k_x$. Potential parameters are $U_0=50E_R$, 
$\gamma=0.98$ and $\phi=\pi/25$. 
\label{fig:psiqubit}}
\end{figure}

To help the qubit addressability, the lattice of rings can be produced by using a $nS\rightarrow nP$ transition; 
then, the individual addressing of lattice sites is always feasible by using a $nS\rightarrow (n+1)P$ transition which has a higher frequency and thus a smaller wavelength. At the same time the spatial stability of the lattice, 
obtained by controlling the phases of the laser fields \cite{Schmiegelow2016}, would ensure the repeatability of 
the addressing. Figure~\ref{fig:psiqubit} shows the spatial and momentum distributions (modulus square of the wave functions) of the two qubit states $|0\rangle$ and $|1\rangle$. Though their spatial densities look similar, we 
observe that the wave functions of the states $|0\rangle$ and $|1\rangle$ are respectively even and odd with respect to $x\to-x$. This means that their Fourier transforms are also respectively even and odd with respect to 
$k_x\to-k_x$. As a consequence, as seen in Fig.~\ref{fig:psiqubit}, states $|0\rangle$ and $|1\rangle$ are easily distinguishable by their momentum distributions, allowing qubit state discrimination for quantum processing via 
time-of-flight measurements. As a specific protocol to achieve the goal, one could selectively excite the atoms in 
a given ring to a hyperfine state which is not trapped by the lattice laser beams. Then, relying on the clearly different momentum distributions of the qubit states, the read-out can be carried out on them via time-of-flight measurements.
 
In our proposed architecture, the coupling between the qubits could be achieved by superposing 
a tailored hologram, generated by a SLM and a high-power objective, to the lattice. 
The idea is to mimic the inductive coupling between superconducting flux 
qubits \cite{Mooij,superc_JJ_Q_bits}. Starting from a lattice at unit filling, the barrier between two adjacent rings would be lowered during a certain coupling time, enhancing (virtual) particle hopping, while being kept 
large enough so that on-site interactions still favor single site occupancy. In this case, the coupling between the two adjacent qubits would be controlled by the ratio between the tunneling rate, the atom-atom interaction strength and the coupling time.

Another promising possibility would be to work with a magnified system (obtained for instance by SLM methods) 
such that a large number of interacting particles can be confined inside each ring. This would provide a platform for a lattice of ring condensates where quantum phase slip tunneling can occur. \cite{qps_mateev,qps_Frank,qps_exp1,qps_exp2,solenov_qubit}. Quantum phase slip is a collective process implying the tunneling of the phase degree of freedom of the cold atoms flowing into adjacent qubits of the lattice.
Such processes occur close to the Mott insulating states in which phase fluctuations are sufficiently strong to trigger the tunneling of the phase states \cite{Nazarov}.
Because of such tunneling, each persistent current will be coupled to the other persistent 
current flowing in the other qubit.
Referring to the superconducting platform, the experiments conducted on circuits 
involving fluxonium architectures evidence quantum phase slip tunneling rates of 
the order of $1-10$ GHz \cite{qps_exp1,Devoret}. With cold atoms however, there has 
been no experiment so far. Clearly, the time scales are very different (milliseconds) 
and therefore quantum phase slips in our cold atom system may be expected in the kHz range. 
A description of protocols based on quantum phase slip and analysis of their performances 
would require a detailed study on its own and is beyond the scope of this paper.
 
Regardless of the actual coupling scenario, effective coupling terms of the form 
$\sigma_z\otimes\sigma_z$ and $\sigma_x\otimes\sigma_x$ between adjacent qubits are expected. 
Then one could envisage implementing a two-qubit controlled-NOT gate analogously to superconducting flux qubits.
It is also worth mentioning that the main source of decoherence in our system is expected to come from collisions with the background gas, leading to decoherence time scales of the order of tens of seconds.
Together with single gate operations, such a system of ring condensates, arranged in a triangular lattice, would have the potential to generate a two-qubit  universal quantum gate 
\cite{superc_JJ_Q_bits, SWAP, loss}. With this approach one could even couple many different pairs of adjacent qubits in parallel. Here again, the spatial stability of the lattice potential is essential for a successful implementation of the scheme.

\section{Conclusion}
\label{sum}
We have proposed a laser scheme providing a possible scalable architecture of ring qubits 
placed in the elementary cells of a triangular lattice and realizing an atomtronic light circuit. 
Each qubit is rendered by a quantum particle moving in the (distorted) ring-shaped minimum of a 
Mexican hat potential. 
The typical spatial extension of each qubit is of a few microns but could be magnified to larger 
sizes by SLM techniques. The obtained triangular 2D array of atomtronic ring qubits can be manipulated 
similarly to superconducting flux qubits, but with an effective magnetic field generated by 
Laguerre-Gauss laser beams imprinting a synthetic gauge field on the atoms. The flux state can 
be determined by interference measurements \cite{eckel2014} or by Doppler measurement of the 
quantized flow state \cite{Kumar2016}. Future studies should consider the role of atom-atom 
interactions and address the coupling between the condensates wavefunctions within adjacent 
ring-shaped potential minima \cite{footnote} as well as the performances of such a system for 
quantum information processing purposes.
Finally, we observe that, beyond quantum information purposes, our scheme could be viewed as 
a quantum simulator made of ultracold atoms vortex arrays \cite{Vignolo07,Burkov06,QuantumSimul} 
or as a quantum sensor based on light-matter angular momentum transfer \cite{Thiel16}.

\section{Acknowledgements}
This research is supported by the National Research Foundation, Prime Minister's Office, Singapore 
and the Ministry of Education-Singapore under the Research Centres of Excellence programme and 
Academic Research Fund Tier 2 (Grant No. MOE2015-T2-1-101). 

\bibliographystyle{apsrev4-1}



\end{document}